\documentclass[conference]{IEEEtran}
\IEEEoverridecommandlockouts
\usepackage{cite}
\usepackage{amsmath,amssymb,amsfonts}
\usepackage{algorithmic}
\usepackage{graphicx}
\usepackage{textcomp}
\usepackage{xcolor}
\usepackage{tabularx}
\usepackage{multirow}

\usepackage{mathtools}
\DeclarePairedDelimiter\ceil{\lceil}{\rceil}

\def\BibTeX{{\rm B\kern-.05em{\sc i\kern-.025em b}\kern-.08em
    T\kern-.1667em\lower.7ex\hbox{E}\kern-.125emX}}
\begin{document}

\title{High-Throughput Split-Tree Architecture for Nonbinary SCL Polar Decoder}

\author{\IEEEauthorblockN{Yaoyu Tao, Cedric Choi}
\IEEEauthorblockA{Qualcomm Wireless R\&D, San Jose, CA 95110 \\
\{yaoyut, cedricc\}@qti.qualcomm.com}
}

\maketitle

\begin{abstract}
Nonbinary polar codes defined over Galois field GF($q$) have shown improved error-correction performance than binary polar codes using successive-cancellation list (SCL) decoding. However, nonbinary operations are complex and a direct-mapped decoder results in a low throughput, representing difficulties for practical adoptions. In this work, we develop, to the best of our knowledge, the first hardware implementation for nonbinary SCL polar decoding. We present a high-throughput decoder architecture using a split-tree algorithm. The sub-trees are decoded in parallel by smaller sub-decoders with a reconciliation stage to maintain constraints between sub-trees. A skimming algorithm is proposed to reduce the reconciliation complexity for further improved throughput. The split-tree nonbinary SCL (S-NBSCL) polar decoder is prototyped using a 28nm CMOS technology for a (128,64) polar code over GF(256). The decoder delivers 26.1 Mb/s throughput, 11.65 Mb/s/mm$^2$ area efficiency and 28.8 nJ/b energy efficiency, outperforming the direct-mapped decoder by 10.3$\times$, 4.4$\times$ and 2.7$\times$, respectively, while achieving excellent error-correction performance. 
\end{abstract}


\section{Introduction}

Binary polar codes \cite{polar_arikan_first} have been adopted in modern communication systems such as fifth generation (5G) wireless communication. However, the error-correction performance is not very competitive with moderate code length. Recently, nonbinary polar codes designed in Galois field GF($q$) (where $q$ is the GF order) have shown improved error-correction performance even with a moderate code length \cite{tal2012constructing,yuan2018construction,mori2010non, chen2018new,cayci2019nonbinary,chen2018two,park2013,feng2020}. Similar to binary polar codes, nonbinary polar codes can be decoded on a trellis using successive cancellation list (SCL) \cite{scl_tal_first} decoder. The computations in nonbinary SCL (NBSCL) decoder are based on log-likelihood-ratio vectors (LLRVs) of length $q$. The error-correction performance can be improved by increasing the GF order $q$ or list size $L$, but the decoding complexity also grows super-linearly. It is crucial yet challenging to design a high-throughput decoder architecture for practical adoption of nonbinary polar codes. 

A variety of speedup techniques have been researched for binary SCL decoding, such as tree pruning \cite{chen2015reduce,chen2013improved,hashemi2016simplified} and tree splitting \cite{li2013parallel,tao2020configurable, shi2019reduced,li2015decision}. Tree pruning reduces the complexity by removing branches with vanishing likelihoods. However, the decoding depth may stay the same, making pruning less effective in reducing the latency. Tree splitting divides the decoding tree to $M$ sub-trees and reduces the decoding depth by $M\times$ theoretically. In practice, the latency improvement is lower than $M \times$, because an extra reconciliation stage is needed to maintain constraints between sub-trees. 

To the best of our knowledge, there's no prior work on hardware implementation of NBSCL decoder. In this work, we apply tree splitting to nonbinary polar codes for improved decoding throughput. Parallel sub-decoders are designed to efficiently support computations in Galois field based on LLRVs. A nonbinary reconciliation processor is developed using a skimming algorithm to minimize the reconciliation latency. The split-tree nonbinary SCL (S-NBSCL) polar decoder is implemented using a 28nm CMOS technology for a (128,64) GF(256) polar code. With list size 4 and split factor 2, the prototype decoder delivers 26.1 Mb/s throughput, 11.65 Mb/s/mm$^2$ area efficiency and 28.8 nJ/b power efficiency while achieving excellent error-correction performance.

\section{Background}
\label{nb_scl_algorithm}

A ($N$,$K$) nonbinary polar code over GF($q$) ($q = 2^r$, $r \in Z^{+}$) has a code length of $N$ symbols, among which $K$ symbols are free symbols that can be used to carry information and the rest $N-K$ symbols are frozen to predetermined GF values. The encoding process can be described as \eqref{f_eq}:

\begin{equation}
\label{f_eq}
c_{0}^{N-1} = u_{0}^{N-1}F^{\otimes n}\text{, } F = \begin{bmatrix} 1 & 0 \\ \alpha & \beta \end{bmatrix}\text{, } \alpha, \beta \in \text{GF}(q)
\end{equation}

\noindent where $u_{0}^{N-1}$ and $c_{0}^{N-1}$ denote the input symbols and the codeword symbols of length $N$, respectively, and $(\cdot)^{\otimes n}$ denotes the $n$-order Kronecker power with $n = \text{log}_2N$. Here $\alpha$ and $\beta$ can be selected for strongest polarization \cite{yuan2018construction} for target code.

\subsection{Nonbinary SCL Polar Decoding}
\label{nb_scl}

\begin{figure}
\centering
\includegraphics[width=\linewidth]{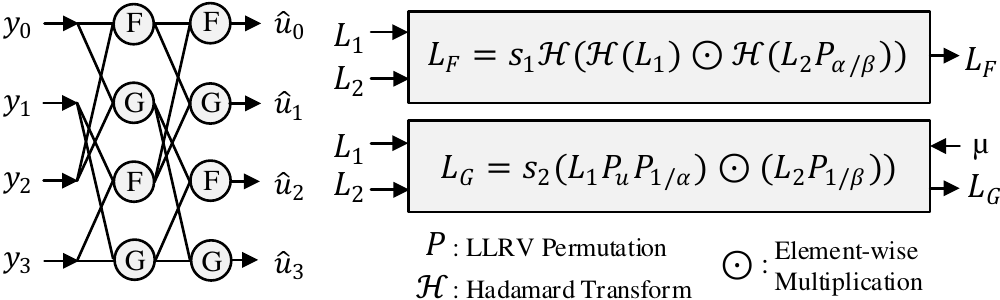}
\caption{NBSCL decoding trellis.}
\label{nb_sc_trellis}
\end{figure}

Nonbinary polar code can be decoded on a trellis using SCL algorithm. \figurename~\ref{nb_sc_trellis} shows an example for $N = 4$. The LLRVs $y_0^{N-1}$ are fed into the trellis from the left hand side and the symbol decisions $\hat{u}_0^{N-1}$ are made on the right hand side. The decoding trellis consists of $n$ stages of F and G functions. The F function receives two LLRVs $L_1$ and $L_2$ and computes the output LLRV $L_{F}$ with a scaling factor $s_1$. The $F$ function needs Hadamard transform (denoted by $\mathcal{H}$) and element-wise multiplication (denoted by $\odot$). The G function receives the partial sum $\mu$ of previously decoded symbols, in addition to the two LLRVs $L_1$ and $L_2$, and computes the output LLRV $L_{G}$ with a scaling factor $s_2$. The permutations (denoted by $P$) are determined by $\alpha$ and $\beta$ selections.

SCL decoding follows a symbol-by-symbol sequential order ($\hat{u}_0 \rightarrow \hat{u}_{N-1}$). For completeness, we briefly introduce the three steps for decoding the $i-$th symbol: 

\noindent\makebox[\linewidth]{\rule{\linewidth}{0.5pt}}

\noindent 1) Selected F and G functions are enabled to compute the trellis output LLRV $L(\hat{u}_i)$. The selections vary symbol to symbol and are predetermined by code's construction; 

\vspace{0.1cm}

\noindent 2) Path metrics (PMs) $P(\hat{u}_{i})$ for the $i$-th symbol are calculated based on $P(\hat{u}_{i-1}$) from the $(i$-$1)$-th symbol and the trellis output LLRV $L(\hat{u}_i)$. If $\hat{u}_i$ is a free symbol, each of the $L$ survival paths from the $(i$-$1)$-th symbol branches to up to $q$ candidate paths; otherwise, each survival path only branches to 1 path based on the pre-determined frozen symbol;

\vspace{0.1cm}

\noindent 3) if $\hat{u}_i$ is a free symbol, up to $qL$ candidate paths are sorted based on their PMs and the top $L$ paths are kept.

\noindent\makebox[\linewidth]{\rule{\linewidth}{0.5pt}}

\noindent The procedure repeats itself until reaching the last symbol. Note that all three steps are needed to decode a free symbol, while step (3) can be by-passed when decoding a frozen symbol.

\subsection{Decoder Design Challenges}

\begin{figure}
\centering
\includegraphics[width=0.90\linewidth]{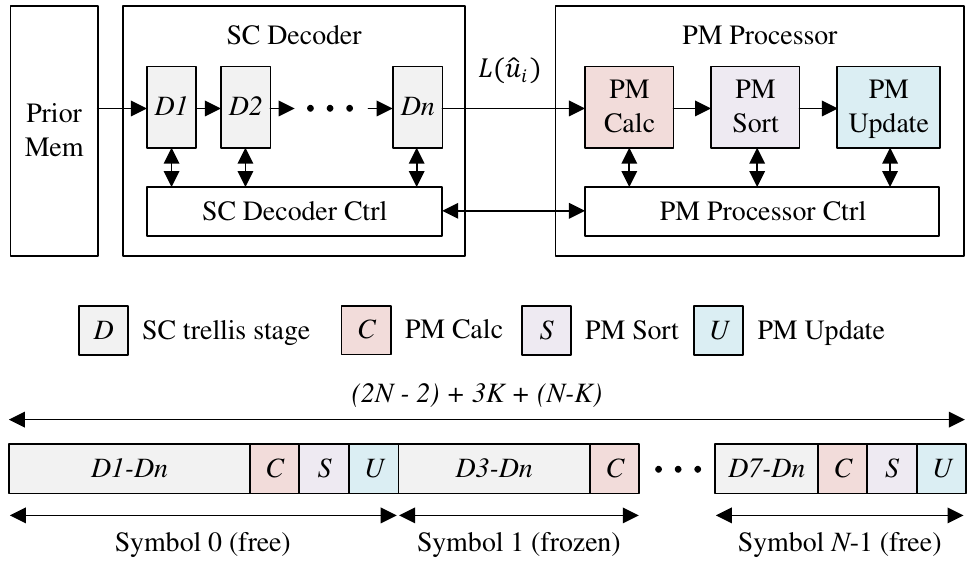}
\caption{Direct-mapped NBSCL decoder architecture and timing scheduling for a ($N$,$K$) nonbinary polar code.}
\label{nb_scl_block_diagram}
\end{figure}

A direct-mapped NBSCL decoder architecture and its processing schedule are shown in \figurename~\ref{nb_scl_block_diagram}. The architecture consists of a SC decoder for trellis computations and a PM processor to maintain $L$ candidate paths. The SC decoder can be pipelined into $n$ stages ($D1$ to $Dn$); however, the decoding latency varies from symbol to symbol because decoding of different symbols uses different F and G functions on the trellis. Suppose it takes 1 unit time for a F or G function. The SC decoder latency per symbol can vary from 1 time unit to $n$ time units. For a codeword of $N$ symbols, the total latency for trellis computations is $2N-2$ time units.

Once the SC decoder stage is completed, the PM processor uses the trellis output LLRV $L(\hat{u}_i)$ to compute the PMs of candidate paths. PMs for candidate paths are sorted and only the top $L$ paths with highest PMs are kept. Suppose PM calculation, PM sort and PM update each take 1 unit time. PM sort and PM update are only incurred if a symbol is a free symbol. For an ($N$,$K$) nonbinary polar code, the direct-mapped NBSCL decoder incurs a latency of $(2N - 2) + 3K + (N-K)$ time units. The decoding speed can be further degraded with larger list size or GF order. For example, the unit time for PM sort grows super-linearly when list size $L$ or GF order $q$ increases. The challenges call for new algorithm and hardware architecture to enable faster decoding.

\section{Algorithm Design for High Throughput}

In this section, we propose tree splitting for SC decoder and skimming for PM processor for improved decoding speed.

\subsection{Nonbinary Split-Tree Decoding}

We first extend split-tree decoding for nonbinary polar codes: an $N$-symbol code can be split into $M$ subcodes of $N/M$ symbols linked by a constraint equation. Frozen symbols on each sub-code are determined by the original $N$-symbol code. The decoding of each subcode is based on a smaller $N/M$-symbol trellis. The proposed S-NBSCL decoder consists of $M$ sub-decoders for $N/M$ symbols that operate in parallel. For the $i$-th symbol ($i = 0 \rightarrow N/M-1$) in each sub-decoder, the decoding has two stages: 

\noindent\makebox[\linewidth]{\rule{\linewidth}{0.5pt}}

\noindent 1) Sub-decoding: $M$ SC sub-decoders operate on their $N/M$-symbol subcode trellises. The $j$-th sub-decoder computes the LLRV of symbol $\hat{u}_{jN/M+i}$ where $j = 0 \rightarrow M-1$. The PMs of candidate paths per sub-decoder (called sub-paths) are computed. If $\hat{u}_{jN/M+i}$ is a free symbol, the $j$-th sub-decoder generates up to $qL$ sub-paths; otherwise, the $j$-th sub-decoder generates $L$ sub-paths. 

\vspace{0.1cm}

\noindent 2) Reconciliation: Global paths are computed based on sub-paths and sorted. The sub-paths that form the top $L$ global paths are distributed to the $M$ sub-decoders.

\noindent\makebox[\linewidth]{\rule{\linewidth}{0.5pt}}

\subsection{Sub-path Skimming for Nonbinary Reconciliation}

In stage 2) of S-NBSCL decoding, if the $M$ symbols on sub-decoders are all free symbols, there could be up to $(qL)^M$ possible global paths made by combinations of sub-paths. However, some global paths are invalid for global PM computation and sorting. For example, if $\hat{u}_0$ is a frozen symbol that is set to a predetermined $\theta \in \text{GF}(q)$, only paths with $\hat{u}_0 = \theta$ are valid. In the best case, only $L^M$ global paths are valid if the $M$ symbols on sub-decoders are all frozen symbols. We observe that sub-paths with smallest PMs have very little effect on the top $L$ global paths. Therefore, we propose a skimming algorithm that drops the sub-paths with small PMs, i.e., only the top $L_s$ sub-paths are kept in each sub-decoder. The reconciliation with skimming has two stages:

\noindent\makebox[\linewidth]{\rule{\linewidth}{0.5pt}}

\noindent 1) Sub-path skimming: The sub-path PMs are locally sorted based on their PMs and only the top $L_s$ sub-paths are kept;

\vspace{0.1cm}

\noindent 2) Global path processing: The $L_s$ sub-paths from each sub-decoder are assembled for $L_s^M$ global paths and sorted. The top $L$ global paths are disassembled and the corresponding sub-paths are distributed to the $M$ sub-decoders.

\noindent\makebox[\linewidth]{\rule{\linewidth}{0.5pt}}

\noindent Performance impact of proposed sub-path skimming is evaluated in Section~\ref{sec:evaluation}. 

\section{High-Throughput Architecture for S-NBSCL Polar Decoding}

\begin{figure}
\centering
\includegraphics[width=.95\linewidth]{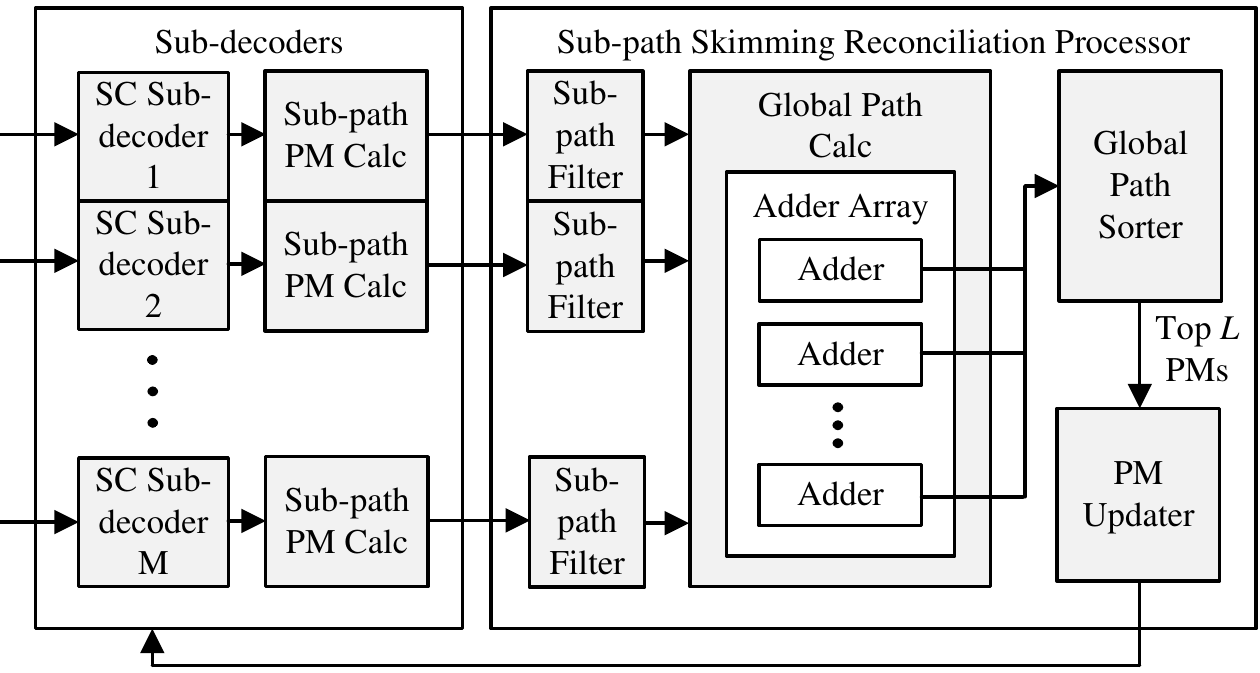}
\caption{Top-level architecture of proposed S-NBSCL decoder.}
\label{nbscl_decoder}
\end{figure}

\begin{figure}
\centering
\includegraphics[width=.95\linewidth]{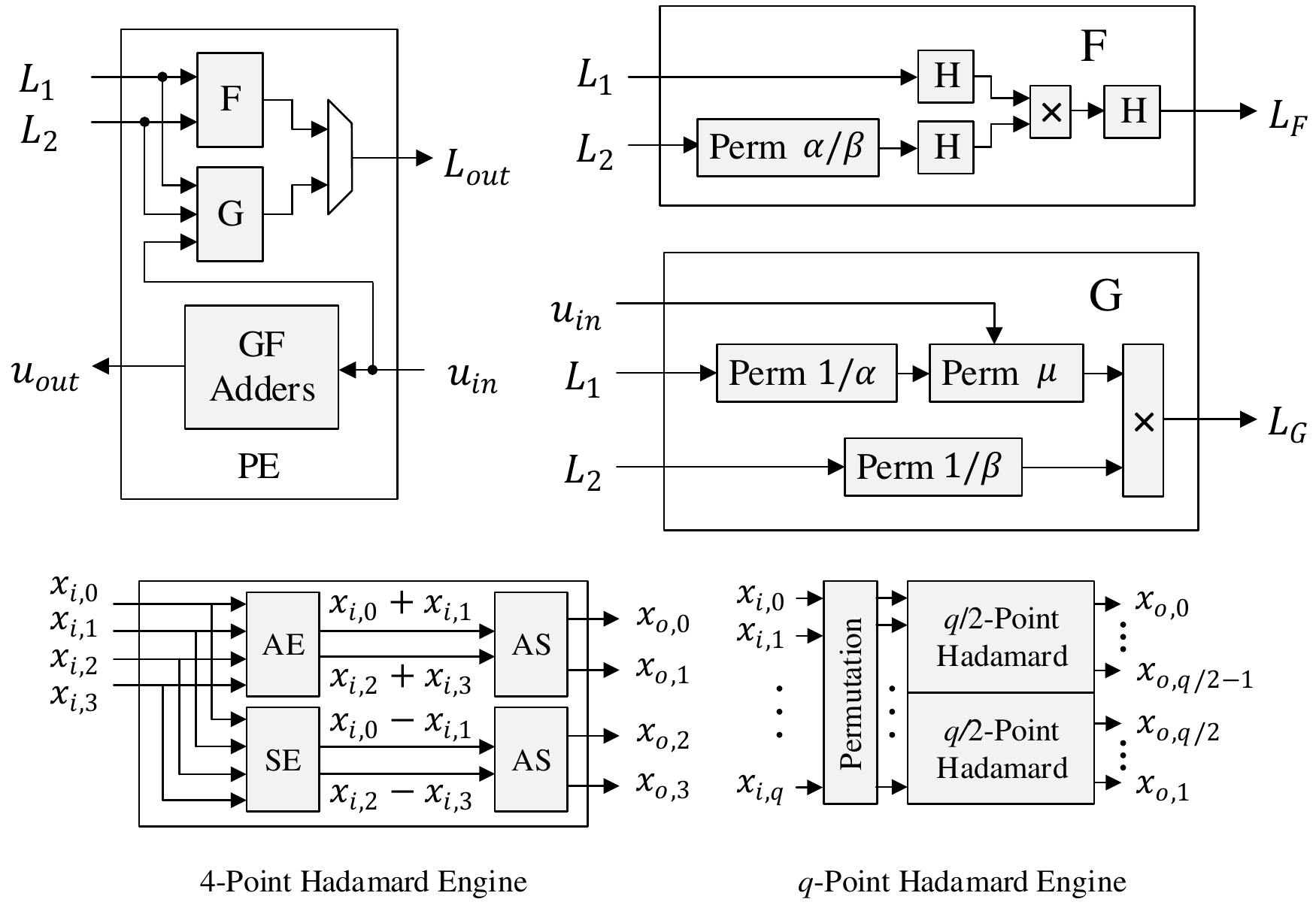}
\caption{Nonbinary PE design in sub-decoders.}
\label{nonbinary_pe}
\end{figure}

\figurename~\ref{nbscl_decoder} presents the S-NBSCL decoder architecture with a split factor $M$. The input LLRVs are equally split to $M$ groups and fed into $M$ parallel sub-decoders. We present the details of sub-decoders and reconciliation processor as below.

\subsection{Sub-decoder Design}

Each sub-decoder consists of a SC sub-decoder of $n_s = \text{log}_2(N/M)$ stages and a sub-path PM calculator. The SC sub-decoder follows the architecture in \figurename~\ref{nb_scl_block_diagram} and the $i$-th stage consists of $2^{n_s-1-i}$ processing elements (PEs). Each stage contains likelihood registers for LLRVs and state registers to store the partial sums of decoded symbols. Each PE consists of a F function, a G function and GF adders (implemented by XORs) as shown in \figurename~\ref{nonbinary_pe}. The LLRV permutation takes 2 cycles, one for look-up-table (LUT) access and one for shifting. The Hadamard transform is implemented using a $q$-input Hadamard engine (denoted by H in \figurename~\ref{nb_scl_block_diagram}); it can be built from 2 $q/2$-input Hadamard engines with an input permutation stage, as shown in \figurename~\ref{nonbinary_pe}. In our design, we pipeline the $q$-input Hadamard engine into $\text{log}_2q$ stages. The $F$ function has a latency of $2\text{log}_2q+3$ cycles and the $G$ function has a latency of 5 cycles. 

In each sub-decoder, a sub-path PM calculator computes the PMs for up to $qL$ sub-paths using an array of $qL$ 2-input adders. These PMs are sent to reconciliation processor for global path assembling.

\begin{figure}
\centering
\includegraphics[width=.95\linewidth]{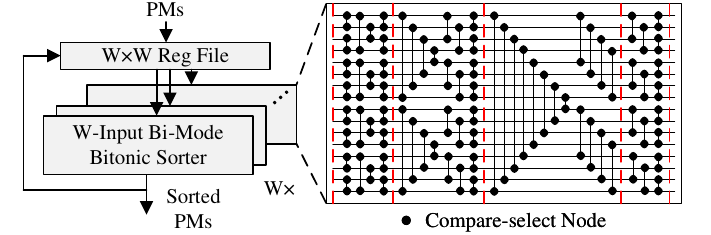}
\caption{Multi-dimensional PM sorter with $W = 16$.}
\label{recon_sorter}
\end{figure}

\subsection{Reconciliation Processor Design}
\label{recon_sorter_sec}

The reconciliation processor consists of 4 stages as shown in \figurename~\ref{nbscl_decoder}. 1) Sub-path filter: for each sub-decoder, the sub-path PMs are filtered based on frozen symbol information and skimming factor $L_s$. A length-$qL$ sorter sorts the sub-path PMs and only the top $L_s$ valid sub-paths are kept; 2) Global path calculator: the valid sub-paths are assembled for up to $L_s^M$ global paths and their PMs are calculated by an array of $L_s^M$ $M$-input adders; 3) Global path sorting: the global paths are sorted based on their PMs using a sorter of length $L_s^M$ and the top $L$ global paths are kept; 4) Path metric update: the top $L$ global paths are disassembled and distributed back to the $M$ sub-decoders, based on which the partial sums in each sub-decoder are updated.

PM sorting can be a major bottleneck in reconciliation. We take the global PM sorting with $M = 2$ and $L_s = 16$ as an example, where a total of $L_s^M = 256$ global PMs need to be sorted. A conventional parallel merge sorter \cite{song2016parallel} takes $L_s^M\text{log}(L_s^M) \approx 616$ cycles. To reduce the sorting latency, we design a 2-dimensional (2D) sorter based on bitonic networks \cite{song2016parallel, mashimo2017high,saitoh2018very,norollah2019rths} as shown in \figurename~\ref{recon_sorter}. The 256 PMs are reshaped into a $W\times W$ matrix where $W = \ceil{\sqrt{L_s^M}} = 16$. The 2D sorter consists of a $W\times W$ register file and $W$ copies of $W$-input bi-mode bitonic networks that can be pipelined into $\text{log}_2W$ stages. The 2D sorting completes in only 6 phases \cite{norollah2019rths} and each phase takes $W+\text{log}_2W = 20$ cycles. It takes only 120 cycles to sort 256 global PMs.

\section{Implementation and Evaluation}
\label{sec:evaluation}

The proposed architecture has been implemented for a (128,64) polar code over GF(256). The decoder utilizes a 8-bit precision for LLRVs and 16-bit precision for PMs. 

\subsection{Error-correction Performance}

\begin{figure}
\centering
\includegraphics[width=\linewidth]{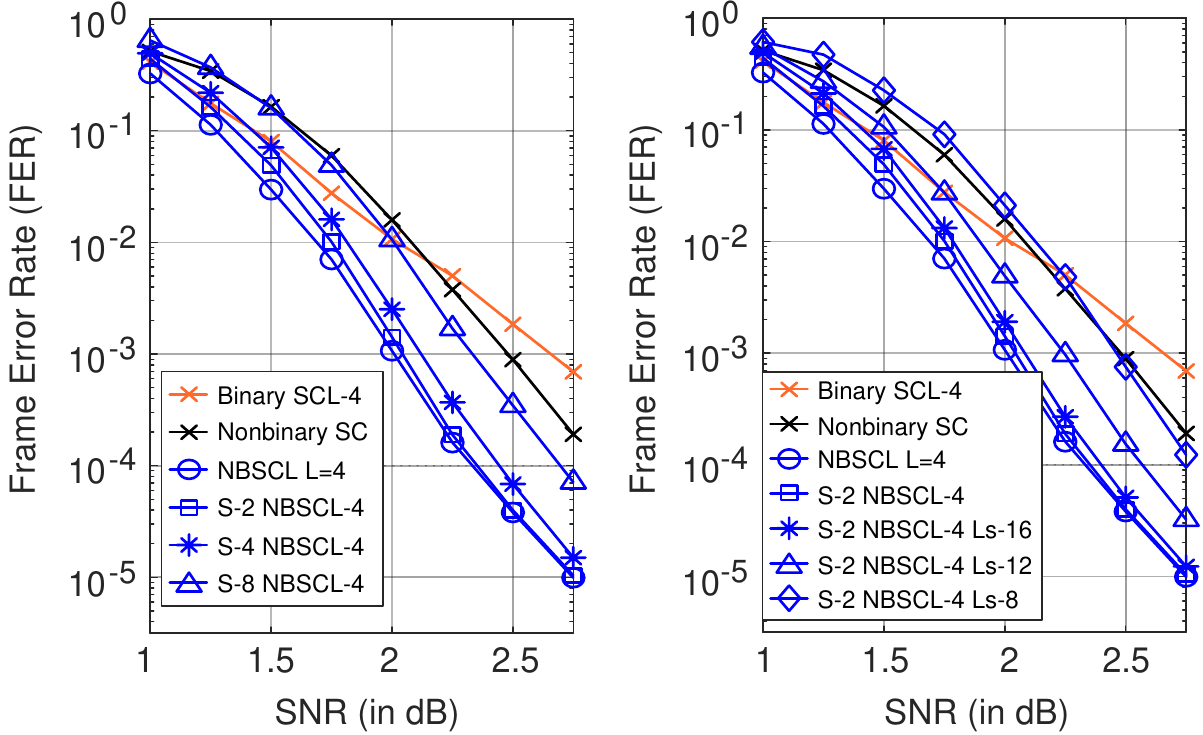}
\caption{Frame error rates of S-NBSCL decoding with sub-path skimming for (128,64) polar code over GF(256).}
\label{fer_fig}
\end{figure}

We evaluate the error-correction perforamnce of proposed split-tree and sub-path skimming algorithms as shown in \figurename~\ref{fer_fig}. We also simulate nonbinary SC for the (128,64) GF(256) polar code and binary SCL $L=4$ for the equivalent 1024-bit polar code for comparison. NBSCL $L=4$ outperforms the binary SCL $L=4$ by 0.63 dB at FER $10^{-3}$. With a split factor $M = 4$, the proposed S-NBSCL degrades the error-correction performance by 0.12 dB compared to NBSCL $L=4$ at FER $10^{-3}$. Increasing the split factor to $M = 8$ or above has significant impact on coding gain. Using a split factor $M = 2$ and a sub-path skimming factor $L_s = 16$, the S-NBSCL is able to achieve only 0.08 dB performance loss compared to NBSCL $L=4$ at FER $10^{-3}$. Further reducing the skimming factor $L_s$ incurs noticeable degradation in error-correction performance. We choose split factor $M = 2$, skimming factor $L_s = 16$ and list size $L = 4$ for our prototype decoder designs. 

\subsection{Hardware Prototyping}

We first implement the direct-mapped NBSCL (DM-NBSCL) decoder in \figurename~\ref{nb_scl_block_diagram} for the (128,64) GF(256) polar code as the baseline. The PE latency is $2\text{log}_2q+3 = 19$ cycles. The SC decoding takes $(2N-2)\times (2\text{log}_2q+3)$ = 4826 cycles. The PM calculator has a 1 cycle latency when using $qL = 1024$ parallel 2-input adders. A conventional parallel merge sorter is implemented to sort the 1024 sub-path PMs in 3083 cycles. The PM updater takes $n = 7$ cycles to back propagate the partial sums of decoded symbols. The total latency for PM processor is $1+3083+7 = 3091$ cycles. The overall latency to decode a frame using direct-mapped decoder is $4826+64+64\times3091 = 202714$ cycles, equivalent to 2.53 Mb/s with 500MHz clock frequency.

The S-NBSCL decoder consists of 2 sub-decoders of 64 symbols and a reconciliation processor with skimming factor $L_s = 16$. Since the latency for sub-decoder is $(2N/M-2)\times (2\text{log}_2q+3)$, decoding a 64-symbol subcode requires 2394 cycles. The reconciliation latency depends on the frozen symbol selections when constructing the polar code. Among 64 decoding levels, 19 levels involve 2 symbols that are all frozen symbols at each sub-decoder and the reconciliation can be by-passed. The remaining 45 levels involve at least one free symbol and requires reconciliation.

\begin{table}
	\centering
	\caption{Implementation summary based on synthesis in 28nm CMOS technology and Ansys PowerArtist simulations}
	\renewcommand{\arraystretch}{1.2}
	\begin{tabular}{|c|c|c|c|c|c|}
        \hline
		\multirow{2}{*}{Decoder} & Thpt. & Area & Area Eff. & Power & Power Eff. \\ 
		 & (Mb/s) & (mm$^2$) & (Mb/s/mm$^2$) & (mW) & (nJ/b) \\ \hline
		 \hline
		 DM-NBSCL & 2.53 & 0.96 & 2.64 & 197.4 & 78 \\ \hline
		 S-NBSCL & 26.1 & 2.24 & 11.65 & 751.1 & 28.8 \\ \hline
	\end{tabular}
	\label{comparison_table}
\end{table}

Inside reconciliation processor, each sub-path filter instantiates a 2D sorter of length $qL = 1024$ (i.e., $W = 32$) that takes $6\times(W+\text{log}_2W) = 222$ cycles. Loading sorter register files and validating the PMs based on frozen conditions require another 34 cycles. The global path calculator can produce $L_s^M = 256$ global PMs in 1 cycle, followed by global path sorter that has a sorting latency of 120 cycles as discussed in Section~\ref{recon_sorter_sec}. The PM updater needs $n_s = 6$ cycles to update the partial sums in each SC sub-decoder. The overall latency to decode a frame using proposed S-NBSCL decoder is $2394 + 19 + 45\times(222+34+1+120+6) = 19648$ cycles, equivalent to 26.1 Mb/s with 500MHz clock frequency. 

Decoder prototypes are synthesized at a 500MHz clock using a 28nm CMOS technology. We use Ansys PowerArtist to simulate the power of decoding one frame considering switching activities. The implementation results are summarized in Table~\ref{comparison_table}. The direct-mapped decoder occupies 0.96 mm$^2$ and consumes 197.4mW power, while the S-NBSCL decoder occupies 2.24 mm$^2$ and consumes 751.1 mW. The proposed S-NBSCL decoder demonstrates a 26.1 Mb/s throughput, a 11.65 Mb/s/mm$^2$ area efficiency and a 28.8 nJ/b energy efficiency, outperforming the direct-mapped decoder by 10.3$\times$, 4.4$\times$ and 2.7$\times$, respectively.

\section{Conclusions}

We present, to the best of our knowledge, the first hardware implementation of nonbinary SCL polar decoder. A high-throughput split-tree architecture is developed that allows parallel decoding of sub-trees with a reconciliation stage. A skimming algorithm is proposed to further reduce the complexity of nonbinary reconciliation for higher throughput. Prototype decoder for a (128,64) polar code over GF(256) with split factor 2 and list size 4 is implemented using a 28nm CMOS technology and runs at 500MHz clock. Simulation results show excellent error-correction performance down to very low error rate. The proposed S-NBSCL occupies 2.24 mm$^2$ in area and delivers a 26.1 Mb/s throughput at 500MHz, consuming 28.8 nJ/b. The throughput, area efficiency and energy efficiency are 10.3$\times$, 4.4$\times$ and 2.7$\times$, respectively, than the direct-mapped NBSCL decoder.

\bibliographystyle{IEEEtran}
\bibliography{IEEEabrv,iscas2022}

\end{document}